\documentclass[aps,amsmath, twocolumn]{revtex4}
\usepackage[dvips]{graphicx}
\usepackage{epsfig}
\usepackage{dcolumn}
\usepackage{bm}
\usepackage[dvips]{color}

\begin{document}

\title{Effects of spin non-collinearities in magnetic nanoparticles}
\author{H. Kachkachi}
\email{hamid.kachkachi@physique.uvsq.fr; URL: http://hamid.kachkachi.free.fr}
\affiliation{Groupe d'Etude de la Mati\`ere Condens\'ee, Universit\'e de
Versailles St. Quentin, 45 av. des Etats-Unis, 78035 Versailles, France}
\begin{abstract}
In a many-spin approach that takes account of the internal structure, microscopic interactions and single-site anisotropies, we investigate the effect of spin non-collinearities induced by the boundary and surface anisotropy on the behaviour of individual magnetic nanoparticles.
Through analytical and numerical calculations, we show that there are mainly two regimes separated by some critical value of the surface anisotropy constant $K_s$  which controls the intensity of spin non-collinearities: i) the so called Stoner-Wohlfarth or N\'eel-Brown regime of a macrospin undergoing a coherent switching, ii) the many-spin regime where the strong spin non-collinearities invalidate the coherent mechanism, and where the particle's magnetic state and switching mechanisms can no longer be modeled by a macrospin.
For small-to-intermediate values of $K_s$, and within two models of surface anisotropy (transverse and N\'eel), the behaviour of the nanoparticle can be modeled by that of a macrospin with an effective potential energy containing a uniaxial and cubic anisotropy terms. This effective spin model provides a crossover between the two regimes.
\end{abstract}
\maketitle
\section{Introduction}
\label{intro}
Reducing the size of magnetic nanoparticles in view of room-temperature technological applications costs a high price. Very small particles present two major difficulties: i) they are less stable against thermal fluctuations, and ii) they have an important contribution from the boundary which induces large deviations from the homogeneous (collinear) magnetic state. As an immediate consequence of the latter the magnetic state of the particle cannot be represented by a macroscopic magnetic moment (single-moment approach), and one has to resort to finer (local) approaches involving the atomic magnetic moment right from the beginning. Only then can one distinguish, and thereby estimate the role of, the various environments inside of the particle, and in particular one may assess the effect of the boundary/surface contribution on the static as well as dynamic behaviour of the magnetic particle. This of course assumes that one has adequately adapted the computing methods.

The magnetic structure and switching mechanisms of many-spin particles (MSP) have been studied for a few years now by many authors using numerical methods, such as the solution of Landau-Lifshitz equation and Monte Carlo technique \cite{MSPInvest}. Analytical calculations using the modified spin-wave theory \cite{kacgar01epjb} and the spherical model \cite{kacgar01physa300} have shown that the spin disorder on the boundary is of long range and propagates deep into the particle. Most of the results emphasize the important role of surface anisotropy and boundary effects and the necessity to take account of the internal structure of nanoparticles.
However, the study of the dynamics of nanoparticles in the many-spin approach presents tremendous difficulties related with the analysis of the energyscape (minima, maxima, and saddle points), which is a crucial step in the calculation of the relaxation rate and investigation of the magnetization reversal at finite temperature.
One may then address the question as to whether there exist some cases in which the full-fledged theory that has
been developed for the one-spin-particle (OSP) approach [see \cite{cofkalwal05worldsc} and references therein] can still be used to describe an MSP. In Refs.~\onlinecite{garkac03prl, kacbon06prb, kacgar06prep} we showed analytically and numerically that indeed, for small deviations from the collinear state, we can model the magnetic state and switching mechanisms of an MSP by a macrospin in an effective potential energy containing a uniaxial and cubic anisotropy terms.

In this paper, we will briefly review through some examples that results that show how the surface defects, and the spin non-collinearities they entail, fundamentally alter the magnetic properties of a small ($3-10$ nm) nanoparticle. In particular, we will show that such a nanoparticle behaves according to different regimes with crossovers determined by the underlying material parameters, notably the surface anisotropy (model and intensity). It turns out that in some typical parameters’ ranges, namely weak surface anisotropy, the OSP approach may be rehabilitated provided that an effective picture of the particle is adopted. More precisely, the MSP may be represented by a single (macroscopic) magnetic moment with an effective energy. This simplification provides us with a useful tool for studying the dynamics, and the thermally activated switching of the magnetization, of a small nanoparticle while taking account, though in a phenomenological manner, the effect of its boundary.
\section{Model and notation}
Surface effects are due to the breaking of crystal-field symmetry at the boundary of the nanoparticle. In order to study
such effects, one has to resort to microscopic theories capable of distinguishing between different atomic environments and taking account of physical parameters, such as surface anisotropy, exchange and dipole-dipole interactions, in addition, of course, to the magneto-crystalline anisotropy in the core and magnetic field.

Our targeted system is an MSP of $\mathcal{N}$ (reduced) atomic magnetic moments $\mathbf{m}_{i}$ with $|\mathbf{m}_{i}|=1$.
The particle's energy includes the exchange, the Zeeman, and anisotropy contributions
\begin{equation}\label{MSPHamiltonian}
\mathcal{H} = -\sum\limits_{i=1}^{\mathcal{N}}
\left[
\mathbf{H}\cdot\mathbf{m}_{i} + K_{i}\,A(\mathbf{m}_{i})
+\sum\limits_{j=1}^{z_{i}}\frac{J_{ij}}{2}\,\mathbf{m}_{i}\cdot\mathbf{m}_{j}.
\right] 
\end{equation}
where $A(\mathbf{s}_{i})$ is the anisotropy function whose expression depends on the type of anisotropy, it may be, e.g., of the uniaxial, cubic, or N\'eel's type, that is
\begin{equation} \label{anisotropy_function}
A(\mathbf{m}_{i})=\left\{
\begin{array}{ll}
(\mathbf{m}_{i}\cdot \mathbf{e}_{i})^{2}, & \quad \mathrm{uniaxial} \\
&  \\ 
-\frac{1}{2}\left[ m_{i,x}^{4}+m_{i,y}^{4} + m_{i,z}^{4}\right] , & \quad
\mathrm{cubic} \\ 
&  \\ 
\frac{1}{2}\sum\limits_{j=1}^{z_{i}}(\mathbf{m}_{i}\cdot \mathbf{u}_{ij})^{2}, & \quad \mathrm{N\acute{e}el}.
\end{array}
\right. 
\end{equation}
with $\mathbf{u}_{ij}$ being a unit vector connecting the nearest neighbors $i,j$.
The anisotropy constant $K_{i}$ may be positive or negative and is denoted by $K_{c}$ if the site $i$ is in the core of the particle and $K_{s}$ if it is on the boundary. A spin in the core has its full coordination number while a spin on the boundary lacks some of its neighbors.

In the present work, the anisotropy in the core is taken as uniaxial with the anisotropy axis $\mathbf{e}_{i}$ along the reference $z$ axis.
For surface spins, we take either the (uniaxial) Transverse surface anisotropy (TSA) model with an axis along the radial (i.e., transverse to the cluster surface) direction or the N\'eel surface anisotropy (NSA) model  \cite{NSA, garkac03prl, jametetal}.
We use the more general model of transverse direction given by the gradient [the vector perpendicular to the isotimic surface $\psi=\mathrm{constant}$ defining the shape of the particle, e.g. a sphere or an ellipsoid]. In the case of a spherical particle, the transverse and radial directions coincide, whereas for another geometry such as an ellipsoid they do not [see Ref.~\cite{kacbon06prb} for more details].

In the sequel, we will make use of the dimensionless parameters $k_c\equiv K_c/J, k_s\equiv K_s/J, t\equiv k_BT/J, h\equiv H/(2K_c)$.
\section{Many-spin particles}
\subsection{Effects of non-collinearities}
\begin{figure}[floatfix]
\begin{center}
\includegraphics[width=4cm]{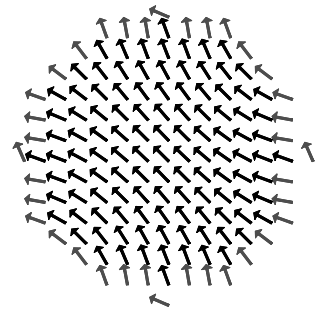}
\end{center}
\caption{ \label{nsa-structure} Magnetic structure of a spherical nanoparticle of linear size $N=15$ with $k_s=2$ for the global magnetization directed along [1,1,0], showing atoms in the plane $z=0$.}
\end{figure}

The $1/D$ surface contribution to $K_{V,{\rm eff}}$, where $D$ is the particle diameter, is in accord with the picture of all magnetic atoms tightly bound by the exchange interaction, whereas only the surface atoms experience the surface anisotropy.
This is definitely true for magnetic films where a huge surface contribution to the effective anisotropy has been observed. The same holds for cobalt nanoclusters of the form of truncated octahedrons \cite{jametetal}, where contributions from different faces, edges, and apices compete resulting in a nonzero, although significantly reduced, surface contribution to $K_{V,{\rm eff}}$. However, for symmetric particle shapes such as cubes or spheres, the symmetry cancels this (first-order) contribution. In this case, one has to take into account deviations from the collinear spin state that result from the competition between the surface anisotropy and the exchange interaction $J$ [see Fig.~\ref{nsa-structure} for the NSA].
%
%
In the case $K_s\gtrsim J$ deviations from collinearity are very strong, and it is difficult, if not impossible, to characterize the particle by a global magnetization suitable for the definition of the effective anisotropy. On the other hand, in the typical case $K_s\ll J$ the magnetic structure is nearly collinear with small deviations that can be computed perturbatively in $K_s/J\ll 1$. The global magnetization vector ${\bf m}_{0}$ can be used to define the anisotropic energy of the whole particle. The key point is that deviations from collinearity and thus the energies of the system are different for different orientations of ${\bf m}_{0},$ even for a particle of a spherical shape, due to  the crystal lattice.

\begin{figure}[ht!]
\includegraphics[width=7.75cm]{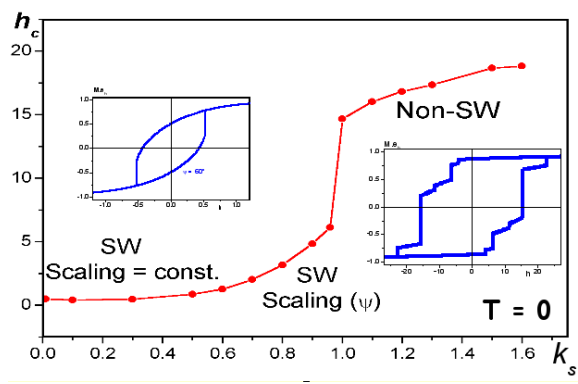}
\caption{Critical field against the ratio of surface anisotropy constant to exchange, $k_s$ for a spherical particle of $360$ spins. The anisotropy is uniaxial in the core (with $k_c = 0.01$) and TSA on the surface.}
\label{SWnonSW}
\end{figure}
%
%
In Fig.~\ref{SWnonSW} we plot the critical field as a function of the ratio of the reduced surface anisotropy to the exchange coupling for a spherical particle of $360$ spins with uniaxial anisotropy in the core and TSA on the surface. Each point on this curve is obtained from the hysteresis loop computed for the corresponding value of $k_s$ (see Ref.~\onlinecite{kacdim02prb_TSANSA} for the details of the procedure).
These results show that upon varying the surface anisotropy one observes a crossover from i) the coherent-reversal regime assuming the particle as a macrospin according to the model of Stoner-Wohlfarth, into ii) the incoherent-reversal regime with cluster-wise switching. In the latter regime, the particle exhibits (due to strong surface anisotropy) new features that are reminiscent of a many-spin system, which cannot be described by a macroscopic approach.
In Refs.~\cite{kacdim02prb_TSANSA} it was shown that for the TSA and NSA there exists a (different) critical value of the surface anisotropy constant that separates the above mentioned regimes.

\begin{figure}[floatfix]
\includegraphics[width=7.75cm]{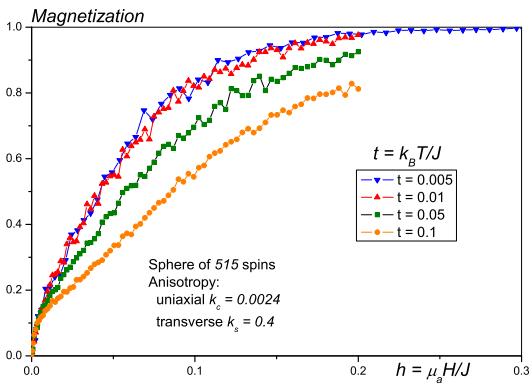}
\caption{Magnetization as a function of field with variable temperature ($t$) for a particle of 515 spins with core uniaxial anisotropy ($k_c = 0.0024$) and transverse surface anisotropy ($k_s = 0.4$).}
\label{SurfaceEffects_MSP}
\end{figure}

Fig.~\ref{SurfaceEffects_MSP} shows the magnetization of a many-spin particle as a function of magnetic field at different temperatures.
%
%
These results, obtained by Monte Carlo simulations (see Ref.~\onlinecite{kacgar01epjb} for details), simply show that,  because of the surface effects, the magnetization does not saturate even at relatively high fields, i.e. $h=0.2$ which corresponds to $H\simeq 16$ Tesla for cobalt particles even at very low temperature, i.e. $t=0.01$ ($T\simeq 1$ K).
\subsection{Crossover to an effective one-spin particle}
With the desire to avoid the difficulties mentioned above, one may address the question as to what extent the dynamics theory developed in the OSP approach \cite{cofkalwal05worldsc} can still be applied to an MSP. However, avoiding somehow the spin non-collinearities induced by surface/interface anisotropy means that some price has to be paid.

Let us now briefly summarize the results obtained in this respect.
Consider a spherical particle with uniaxial anisotropy in the core and TSA on the surface. Then, to compute the contribution of surface anisotropy in Eq.~(\ref{MSPHamiltonian}) we replace the number of nearest neighbors of a surface atom by its average value
\begin{equation}\label{zAvr}
z_{i\alpha }\Rightarrow \overline{z}_{i\alpha }=2-|n_{\alpha }|/\max \left\{|n_{x}|,|n_{y}|,|n_{z}|\right\} .
\end{equation}
where $n_{\alpha }$ is the $\alpha $-component of the normal to the surface ${\bf n}$.
The surface-energy density can then be obtained by multiplying the contribution in Eq.\ (\ref{MSPHamiltonian}) by the surface  atomic density $f({\bf n})=\max \left\{ |n_{x}|,|n_{y}|,|n_{z}|\right\}$:
\begin{equation}\label{ESPerp}
E_{S}({\bf m,n}) = - K_s\,( \mathbf{m}\cdot\mathbf{n})^{2}\,f(\mathbf{n})
\end{equation}
At equilibrium, in the continuous approximation the Landau-Lifshitz equation reduces to
\begin{equation}\label{LLEqEqui}
{\bf m\times H}_{{\rm eff}}=0,\qquad {\bf H}_{{\rm eff}}={\bf H}_{A}+J\Delta
{\bf m,}
\end{equation}
where $\Delta $ is the Laplace operator and the anisotropy field ${\bf H}_{A}$ contains contributions from the core and the surface
\begin{equation}\label{HADelta}
{\bf H}_{A} = -\frac{\delta E_{C}}{\delta{\bf m}} -\frac{\delta E_{S}}{\delta{\bf m}}\delta (r-R),\qquad R\equiv \frac{1}{2}\left( N-1\right) ,
\end{equation}
where $N$ is the side of the cube inside which the particle is cut.

For $K_s\ll J$ the deviations of ${\bf m(r)}$ from the homogeneous state ${\bf m}_{0}$ are small and one can linearize the problem as follows
\begin{eqnarray}\label{mAnsatz}
&&{\bf m(r)} \cong {\bf m}_{0}+{\bf \psi }({\bf r,m}_{0}) = {\bf m}_{0}+{\bf e}_1\psi_1+{\bf e}_2\psi_2,\nonumber\\
&&\psi \equiv |{\bf \psi }|\ll 1.
\end{eqnarray}
The correction ${\bf \psi }$ is the solution of the Helmholtz equation with boundary conditions
\begin{eqnarray}\label{HelmoltzEqts}
&&\left( \Delta -k_{\alpha }^{2}\right) \psi _{\alpha }=-\frac{1}{J}\left( 
\mathbf{H}_{SA}^\bot\cdot \mathbf{e}_{\alpha }\right) ,\quad
\alpha =1,2,  \nonumber\\
&&k_{1}^{2}=\frac{2K_c}{J}\left[ 2m_{0z}^{2}-1\right] ,\quad k_{2}^{2}=\frac{2K_c}{J}m_{0z}^{2},\\
&&\mathbf{H}_{SA}^\bot=\left[ \frac{\delta E_{S}(\mathbf{m}_{0})}{\delta \mathbf{m}}-\left( \frac{\delta E_{S}(\mathbf{m}_{0})}{\delta \mathbf{m}}\cdot \mathbf{m}_{0}\right) \mathbf{m}_{0}\right]\delta(r-R).\nonumber
\end{eqnarray}
where ${\bf n\equiv r/}R$.

The solution $\psi$ of Eq.~(\ref{HelmoltzEqts}) has the form \cite{garkac03prl}
\begin{equation}
{\bf \psi }({\bf r,m})=\frac{1}{4\pi }\int_{S}d^{2}{\bf r}^{\prime }G({\bf r,r}^{\prime }){\bf f(m,n}^{\prime }{\bf )}  \label{psiSol}
\end{equation}
where $G({\bf r,r}^{\prime })$ is the Green function of the problem.

In the absence of core anisotropy, where the Helmholtz equation reduces to a Laplace equation, an extact expression (call it $G^{(0)}({\bf r,r}^{\prime })$) for $G({\bf r,r}^{\prime })$ was obtained in Ref.~\cite{garkac03prl} .
In the presence of core anisotropy we have recently found \cite{kacgar06prep} an approximate expression for the Green's function which contains a correction to $G({\bf r,r}^{\prime })$ of the order $k_\alpha^2\propto K_c/J \ll 1$. This correction is then rewritten as a convolution of two $G^{(0)}$'s.
Collecting all contributions from the core and surface, we showed that the effective energy of an MSP particle, in the absence of magnetic field, is written as
\begin{equation}
\mathcal{E}_{\mathrm{eff.}} = - K_2\,m_{z}^{2}+K_4(m_{x}^{4}+m_{y}^{4}+m_{z}^{4}).
\label{UniaxialCubicEnergy}
\end{equation}
The coefficient $K_2$ of the second-order contribution is in fact the result of two contributions, one coming from the initial core uniaxial anisotropy and a new contribution that is induced by the surface anisotropy. The latter contribution is much smaller than the former because its coefficient contains the product $(K_c/J)(K_s^2/J)\ll 1$. The details of the calculations will be published elsewhere \cite{kacgar06prep}. The $4^{\mathrm{th}}$-order coefficient $K_4$ was found in Ref.~\cite{garkac03prl} to be given by
\begin{equation}\label{Keff}
K_4 = \kappa \frac{\mathcal{N}K_s^2}{z J},
\end{equation}
where $\mathcal{N}, K_s, z, J$ are respectively the number of atoms, the surface anisotropy constant (transverse or N\'eel), the coordination number, and the exchange coupling of the many-spin particle. $\kappa$ is a surface integral that depends on the underlying lattice, the shape, and the size of the particle and also on the surface-anisotropy model. For a spherical particle (of $\sim 1500$ spins) cut from a simple cubic lattice with N\'eel's surface anisotropy, $\kappa\simeq 0.53465$.

In \cite{kacbon06prb} we confirmed this result by numerical calculations of the field behaviour of the net magnetization and effective energyscape.
Before we discuss the results, we briefly explain the method we used to obtain them. Because we are dealing with an MSP, the energyscape cannot be represented in terms of the coordinates of all spins. Instead, we may represent it in terms of the coordinates of the particle's net magnetization. For this purpose, we fix the global or net magnetization, $\mathbf{m}$, of the particle in a desired direction ${\bm m}_{0}$ ($|{\bm m}_{0}|=1$) by using the energy
function with a Lagrange multiplier ${\bm \lambda }$ \cite{garkac03prl}:
\begin{equation}  \label{FFuncDef}
\mathcal{F}=\mathcal{H}-\mathcal{N}{\bm \lambda \cdot }\left( {\bm m}-{\bm m}_{0}\right) ,\qquad {\bm m\equiv }\frac{\sum_{i}\mathbf{s}_{i}}{\left|\sum_{i}\mathbf{s}_{i}\right| }.
\end{equation}
To minimize $\mathcal{F},$ we solve the evolution equations
\begin{eqnarray}
\mathbf{\dot{s}}_{i} &=&-\left[ \mathbf{s}_{i}\times \left[ \mathbf{s}_{i}\times \mathbf{F}_{i}\right] \right] ,\qquad \mathbf{F}_{i}\equiv-\partial \mathcal{F}/\partial \mathbf{s}_{i}  \nonumber  \label{LLEqs} \\
{\dot{\bm \lambda }} &=&\mathbf{\partial }\mathcal{F}/\partial {\bm \lambda=-}\mathcal{N}\left( {\bm m}-{\bm m}_{0}\right) ,
\end{eqnarray}
starting from $\mathbf{s}_{i}={\bm m}_{0}=\mathbf{m}$ and ${\bm\lambda =0,}$ until a stationary state is reached. In this state ${\bm m}={\bm m}_{0}$ and $\left[ \mathbf{s}_{i}\times \mathbf{F}_{i}\right] =0,$ i.e., the torque due to the term $\mathcal{N}{\bm\lambda \cdot }\left( {\bm m}-{\bm m}_{0}\right) $ in $\mathcal{F}$ compensates for the torque acting to rotate the global magnetization towards the minimum-energy directions [see discussion in Ref.~\cite{garkac03prl}]. The orientation of the net magnetization is then given either in Cartesian coordinates $(m_{x},m_{y},m_{z})$ or in spherical coordinates $(\theta_{n},\varphi_{n})$.

\begin{figure}[floatfix]
\includegraphics[width=8cm]{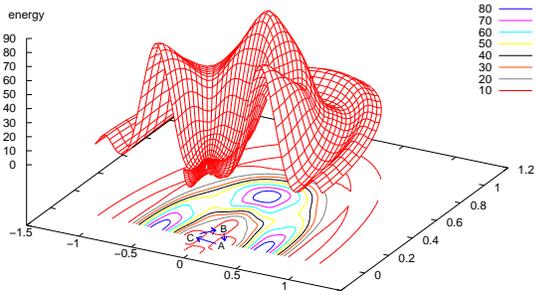}
\caption{Plot of $(\frac{\theta}{\pi}\cos \varphi ,\frac{\theta}{\pi}%
\sin \varphi ,E)$ for the same parameters as before.}
\label{energy_contour_plot}
\end{figure}

\begin{figure}[floatfix]
\includegraphics[width=8cm]{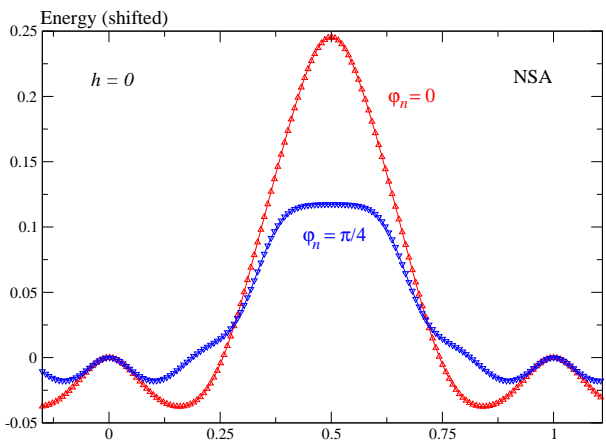}
\caption{Energy as a function of the polar angle $\theta_{n}$ for a spherical particle with uniaxial anisotropy in the core and N\'eel anisotropy on the surface. $\phi_n=0,\pi/4$ and $h=0$.}
\label{NSA360kc001ks03_p0p45}
\end{figure}

We now discuss the main results.
%
\begin{figure*}[floatfix]
\includegraphics[width=4cm, angle=-90]{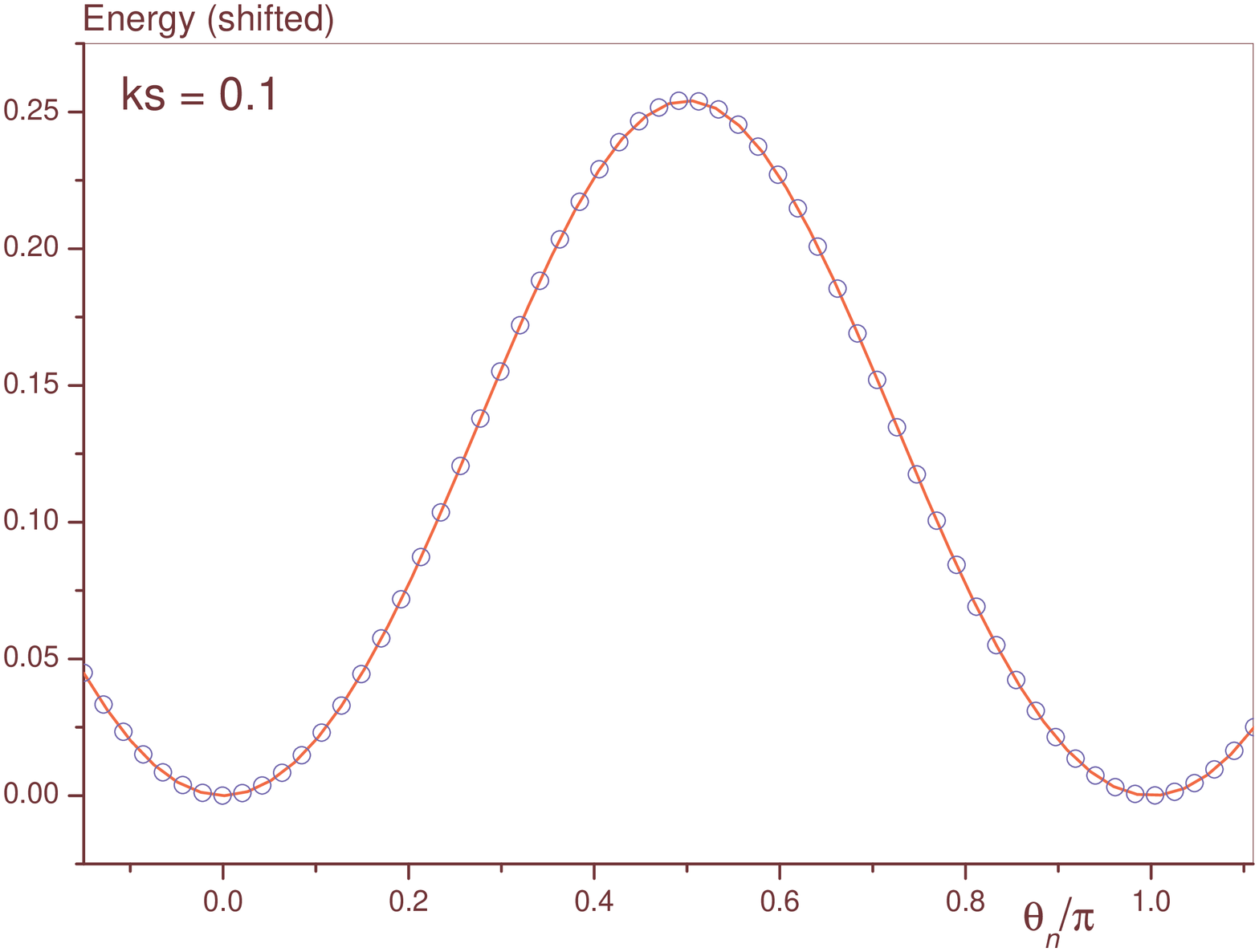}
\includegraphics[width=4cm, angle=-90]{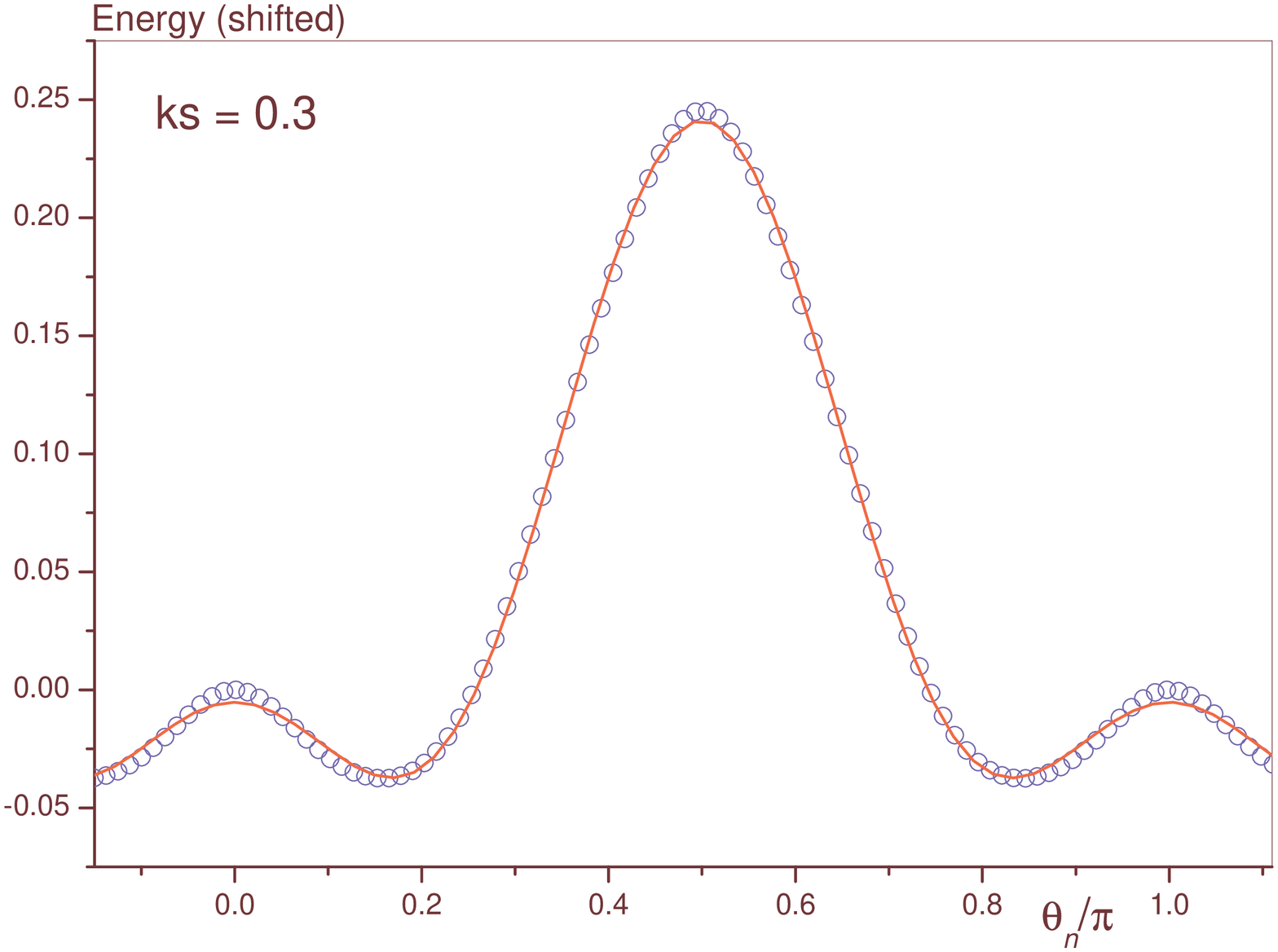}
\includegraphics[width=4cm, angle=-90]{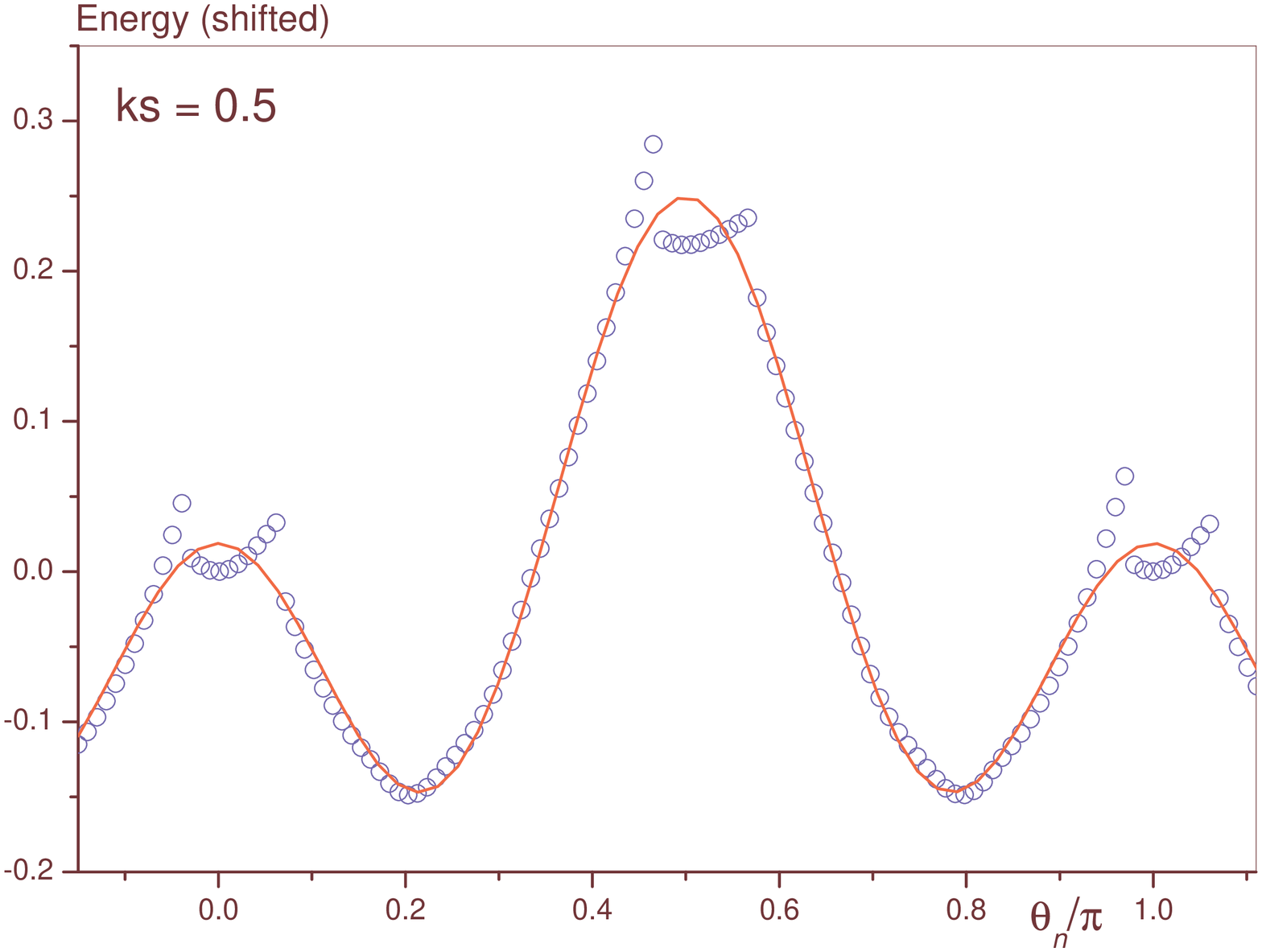}
\caption{$2D$ energyscape for a spherical particle of TSA with varying constant $k_s$. The symbols are results of the numerical calculations for the MSP, and the full lines are fits using Eq. (\ref{UniaxialCubicEnergy}).
$\varphi_n=0$ and $h=0$.}
\label{enscape2D_h0_varyks}
\end{figure*}
%
The first panel of Fig.~\ref{enscape2D_h0_varyks} ($k_s=0.1$) shows that, for very small $k_s$, the energyscape of an MSP is well recovered by the effective energy in Eq.~(\ref{UniaxialCubicEnergy}).
As $k_s$ increases [see middle panel, $k_s=0.3$], some deviations start to be seen, and for relatively larger values of $k_s$ a fit with Eq.~(\ref{UniaxialCubicEnergy}) is no longer possible. In fact, in this regime strong deviations from collinearity develop, especially near maxima and saddle points, as can be seen on the panel with $k_s=0.5$ in Fig.~\ref{enscape2D_h0_varyks}. In fact, in this case the Lagrange-parameter method introduced in \cite{garkac03prl} fails because the magnetic state of an MSP can no longer be represented by a net magnetization.
These results imply that the effect of the spin non-collinearities on the energy is to split the minimum at $\theta_n=0$, defined by the uniaxial anisotropy, into four minima at $\theta_n\sim 28^{\circ }$ and $\varphi_n=0,\pm\pi/2, \pi$, reminiscent of cubic anisotropy [see Fig.~\ref{energy_contour_plot}].
These minima are connected by saddle points at $\varphi_n=\pm \pi /4$ and $\pm 3\pi /4$ and the point at $\theta_n=0$ becomes a small local maximum. The four minima exist over a finite range of the applied field, although their positions change continuously as a function of the field \cite{kacbon06prb}.
%
%
Fig.~\ref{NSA360kc001ks03_p0p45} is a plot of the $2D$ energyscape for a spherical particle with uniaxial anisotropy in the core, as before, but now with NSA on the surface.
%
%
It is clear that the cubic-anisotropy features are also seen in the case of NSA, namely that i) the energy minima are not along the directions ($\theta_{n}=0,\pi$) of the core easy axis; these directions having become local maxima, as discussed in the case of TSA, and ii) there is a clear dependence on the azimuthal angle $\varphi_n$.
In addition, more extensive calculations \cite{fesenkoetal06preprint} have shown that similar features are also observed for other crystal structures (fcc, bcc, etc.).

These results agree and complement those of Ref.~\cite{kacdim02prb_TSANSA}, where it was shown that for the TSA and NSA there exists a (different) critical value of the surface anisotropy constant that separates i) the OSP Stoner-Wohlfarth (SW) regime of coherent switching and ii) the MSP regime where the strong spin non-collinearities invalidate the coherent mechanism, and the particle can no longer be modeled by an effective OSP.
Obviously, for very small surface anisotropy the cubic contribution becomes negligible [see the first panel in Fig.~\ref{enscape2D_h0_varyks}, $k_s=0.1$] and the OSP SW model provides a good approximation to the MSP.
Accordingly, some experimental macroscopic estimations of the surface anisotropy constant yield, e.g., for cobalt $k_s\simeq 0.1$ \cite{skocoe99iop}, for iron $k_s\simeq 0.06$ \cite{urquhartetal88jap}, and for maghemite particles $k_s\simeq 0.04$ \cite{perrai05springer}.
However, one should not forget that this effective constant depends on the particle's size, among other parameters such as the material composition, and for, e.g. a diameter of $2$ nm we may expect stronger anisotropies.
\section{Surface effects on the dynamics}
In the particular (and typical) situation, considered above, where the exchange interaction is much stronger than the anisotropy energy we find that the magnetization can be represented as the spatially homogeneous ``global" magnetization plus a small inhomogeneous contribution that we calculated analytically and numerically \cite{garkac03prl, kacbon06prb, kacgar06prep}.
The latter is induced by surface anisotropy and it is maximal near the surface but can extend deeply into the body of the particle. It describes the adjustment of the magnetization to the conditions at the surface by minimizing the total energy with fixed direction of the global magnetization. As a result, we obtain the effective particle's energy that depends on the orientation of its net magnetization and arises because of surface anisotropy. This contribution is of second order in the surface anisotropy and it adds to other terms, such as the bulk anisotropy and the first-order contribution from the surface anisotropy, which disappear in samples of cubic or spherical shape.
These contributions to the energy of a magnetic nanoparticle are crucial to its dynamical behavior, in particular, in the ferromagnetic resonance (FMR). Accurately taking all of them into account should make it possible to determine the bulk and surface anisotropies from the experimental data. An interesting problem is the dynamical aspect of the magnetization adjustment mentioned above. If the anisotropies are much smaller than exchange interaction, the exchange-driven adjustment is much faster than the global precession of the magnetization induced by the anisotropy. Then, these adjustment modes behave adiabatically at low frequencies and the effective OSP energy is a good approximation.
For materials with a very strong surface anisotropy such separation of dynamical scales is no longer valid, and the dynamics of such nanoparticles becomes an essentially many-body process. FMR experiments on magnetic nanoparticles should allow to estimate the values of the surface anisotropy and detect different regimes of their dynamical behavior.

The effective OSP energy (\ref{UniaxialCubicEnergy}) could be used in investigating the thermally activated reversal of a small nanoparticle. Indeed, this effective energy, together with (\ref{Keff}), allows us to include surface effects, though in an approximate way, while offering a considerable simplification over the initial many-spin system.
However, one should note that the effective potential energy contains a quartic term in the net magnetization components, which renders the analysis of the energyscape somewhat more involved but still tractable \cite{kalmykov00prb}, as opposed to the case of a many-spin particle.
In Ref.~\onlinecite{newell05ggg12} it was shown that for the case of mixed uniaxial and cubic anisotropies, there are two different relaxation rates, one for the parallel component of the magnetization and the other for the component perpendicular to the core uniaxial anisotropy axis. Using Eq.~(\ref{Keff}) one finds, for instance, that the parallel relaxation rate increases with increasing surface anisotropy while the perpendicular component has the opposite behaviour.
This issue requires a more thorough investigation. Nevertheless, for small enough surface anisotropy for which the effective energy (\ref{UniaxialCubicEnergy}) holds best, the minima are mainly defined by the uniaxial anisotropy. The effect of the cubic-anisotropy contribution is then to modify the loci of the saddle points and their number, and thereby the calculation of the relaxation rates becomes more involved \cite{kalmykov00prb}.
\section{Conclusion}
We have studied the properties of magnetic nanoparticles with the many-spin approach including the internal structure and physical parameters such as core and surface anisotropy and exchange interactions.
We have shown that upon varying the surface anisotropy constant there appear three regimes [see, e.g., Fig.~\ref{SWnonSW}]: i) When this constant is very small with respect to the exchange coupling all observables can be scaled recovered within the Stoner-Wohlfarth macroscopic approach. ii) For small and intermediate values, the energy, and thereby the related quantities, of the many-spin particle can be modeled by an effective expression containing a surface-renormalized uniaxial contribution and a $4^{\mathrm{th}}$-order contribution that is induced by the surface anisotropy. iii) For large values of the surface anisotropy constant the spin non-collinearities are so strong that the magnetic state of a nanoparticle can no longer be represented by a net magnetization. In this case, the switching of the particle's magnetization proceeds cluster-wise.

In the intermediate regime, we have also discussed the possibility of studying the effect of surface anisotropy on the relaxation rate and speculated how to determine the nature and intensity of the surface anisotropy using the technique of ferromagnetic resonance.

However, the present models on which all calculations are based suffers from some deficiencies: For instance, it is assumed that the crystal structure on the surface is the same as in the core with the same atomic lattice parameters. This cannot be wholly true considering the possibility of surface reconstructions. Of course, these models do include apices, edges and facets, and the possibility of taking the exchange coupling on the surface as different from that in the core, or that between the core and the surface.
However, there is no guarantee that it will become experimentally possible in the near future to estimate the atomic positions and lattice parameters and may be theoretically possible to perform crystal field calculations, and eventually check these assumptions.
Likewise, the magneto-crystalline anisotropy constant and exchange coupling in the core are taken as those in the bulk underlying material. However, we have shown that the core of a nanoparticle does not enjoy the properties of the underlying bulk material. For a given material such parameters could vary with the radius of the particle. In addition, the intensity and nature of surface anisotropy constitute a real challenge. There are many estimates but no firm understanding is achieved yet as to how surface anisotropy stems from the atomic structure in nanoparticles of reasonable size.

For these interrogations to receive answers experimentalists will have to devise ever more sensitive equipments and measurement techniques in order to probe the intrinsic properties of nanomagnets and have direct access to the related physical observables. This, of course, will require new strategies for further isolating the nanomagnets and rid them off the influence of their hosting matrices and mutual interactions.
On the other hand, understanding the influence on the particles static and dynamic behavior of the surrounding matrix and the inter-particle interactions, is of paramount importance to efficient practical applications.

New experiments on adequately prepared particles are needed to check the validity of these results which, if confirmed,  should help us better understand the dynamics of nanoparticles beyond the macrospin approximation.
%


\begin{thebibliography}{27}
\expandafter\ifx\csname natexlab\endcsname\relax\def\natexlab#1{#1}\fi
\expandafter\ifx\csname bibnamefont\endcsname\relax
  \def\bibnamefont#1{#1}\fi
\expandafter\ifx\csname bibfnamefont\endcsname\relax
  \def\bibfnamefont#1{#1}\fi
\expandafter\ifx\csname citenamefont\endcsname\relax
  \def\citenamefont#1{#1}\fi
\expandafter\ifx\csname url\endcsname\relax
  \def\url#1{\texttt{#1}}\fi
\expandafter\ifx\csname urlprefix\endcsname\relax\def\urlprefix{URL }\fi
\providecommand{\bibinfo}[2]{#2}
\providecommand{\eprint}[2][]{\url{#2}}

\bibitem[{\citenamefont{{H. Kachkachi and M. Dimian}}(2002)}]{kacdim02prb_TSANSA}
\bibinfo{author}{\bibnamefont{{H. Kachkachi and M. Dimian}}},
  \bibinfo{journal}{Phys. Rev. B} \textbf{\bibinfo{volume}{66}},
  \bibinfo{pages}{174419} (\bibinfo{year}{2002});
%
\bibinfo{author}{\bibnamefont{{H. Kachkachi and H. Mahboub}}},
  \bibinfo{journal}{J. Magn. Magn. Mater.} \textbf{\bibinfo{volume}{278}},
  \bibinfo{pages}{334} (\bibinfo{year}{2004}).

\bibitem[{\citenamefont{{D.A. Dimitrov and Wysin}}(1994)}]{MSPInvest}
\bibinfo{author}{\bibnamefont{{D.A. Dimitrov and Wysin}}},
  \bibinfo{journal}{Phys. Rev. B} \textbf{\bibinfo{volume}{50}},
  \bibinfo{pages}{3077} (\bibinfo{year}{1994}); 
%
\bibinfo{author}{\bibnamefont{{H. Kachkachi et al.}}}, \bibinfo{journal}{Eur. Phys. J. B} \textbf{\bibinfo{volume}{14}},
  \bibinfo{pages}{681} (\bibinfo{year}{2000});
%
\bibinfo{author}{\bibnamefont{{O. Iglesias and A. Labarta}}},
  \bibinfo{journal}{Phys. Rev. B} \textbf{\bibinfo{volume}{63}},
  \bibinfo{pages}{184416} (\bibinfo{year}{2001});
%
\bibinfo{author}{\bibnamefont{{Y. Labaye et al.}}}, \bibinfo{journal}{J. Appl. Phys.}
  \textbf{\bibinfo{volume}{91}}, \bibinfo{pages}{8715} (\bibinfo{year}{2002});
%
\bibinfo{author}{\bibnamefont{{E. De Biasi et al.}}}, \bibinfo{journal}{Phys. Rev. B}
  \textbf{\bibinfo{volume}{71}}, \bibinfo{pages}{104408}
  (\bibinfo{year}{2005});
%
\bibinfo{author}{\bibnamefont{{E. Eftaxias and K.N. Trohidou}}},
  \bibinfo{journal}{Phys. Rev. B} \textbf{\bibinfo{volume}{71}},
  \bibinfo{pages}{134406} (\bibinfo{year}{2005});
%
\bibinfo{author}{\bibnamefont{{E. De Biasi et al.}}}, \bibinfo{journal}{Eur. Phys. J. B}
  \textbf{\bibinfo{volume}{51}}, \bibinfo{pages}{65} (\bibinfo{year}{2006}).

\bibitem[{\citenamefont{{H. Kachkachi and D.A. Garanin}}(2001)}]{kacgar01epjb}
\bibinfo{author}{\bibnamefont{{H. Kachkachi and D.A. Garanin}}},
  \bibinfo{journal}{Eur. Phys. J. B} \textbf{\bibinfo{volume}{22}},
  \bibinfo{pages}{291} (\bibinfo{year}{2001}).

\bibitem[{\citenamefont{{H. Kachkachi and D. A.
  Garanin}}(2001)}]{kacgar01physa300}
\bibinfo{author}{\bibnamefont{{H. Kachkachi and D. A. Garanin}}},
  \bibinfo{journal}{Physica A} \textbf{\bibinfo{volume}{300}},
  \bibinfo{pages}{487} (\bibinfo{year}{2001}).

\bibitem[{\citenamefont{{W. T. Coffey, Yu. P. Kalmykov, and J.
  Waldron}}(2005)}]{cofkalwal05worldsc}
\bibinfo{author}{\bibnamefont{{W. T. Coffey, Yu. P. Kalmykov, and J.
  Waldron}}}, \emph{\bibinfo{title}{The Langevin Equation}}
  (\bibinfo{publisher}{World Scientific}, \bibinfo{address}{Singapore},
  \bibinfo{year}{2005}).

\bibitem[{\citenamefont{{D.A. Garanin and H. Kachkachi}}(2003)}]{garkac03prl}
\bibinfo{author}{\bibnamefont{{D.A. Garanin and H. Kachkachi}}},
  \bibinfo{journal}{Phys. Rev. Lett.} \textbf{\bibinfo{volume}{90}},
  \bibinfo{pages}{65504} (\bibinfo{year}{2003}).

\bibitem[{\citenamefont{{H. Kachkachi and E. Bonet}}(2006)}]{kacbon06prb}
\bibinfo{author}{\bibnamefont{{H. Kachkachi and E. Bonet}}},
  \bibinfo{journal}{Phys. Rev. B} \textbf{\bibinfo{volume}{73}},
  \bibinfo{pages}{224402} (\bibinfo{year}{2006}).

\bibitem[{\citenamefont{{H. Kachkachi and D.A. Garanin}}(2006)}]{kacgar06prep}
\bibinfo{author}{\bibnamefont{{H. Kachkachi and D.A. Garanin}}},
  \bibinfo{journal}{In preparation}  (\bibinfo{year}{2006}).

\bibitem[{\citenamefont{{L. N\'eel}}(1954)}]{NSA}
\bibinfo{author}{\bibnamefont{{L. N\'eel}}}, \bibinfo{journal}{J. Phys. Radium}
  \textbf{\bibinfo{volume}{15}}, \bibinfo{pages}{225} (\bibinfo{year}{1954});
%
\bibinfo{author}{\bibnamefont{{R.H. Victora and J.M. McLaren}}},
  \bibinfo{journal}{Phys. Rev. B} \textbf{\bibinfo{volume}{47}},
  \bibinfo{pages}{11583} (\bibinfo{year}{1993});
%
\bibinfo{author}{\bibnamefont{{D.S. Chuang, C.A. Ballantine, and R.C.
  O'Handley}}}, \bibinfo{journal}{Phys. Rev. B} \textbf{\bibinfo{volume}{49}},
  \bibinfo{pages}{15084} (\bibinfo{year}{1994}).

\bibitem[{\citenamefont{{M. Jamet et al.}}(2001)}]{jametetal}
\bibinfo{author}{\bibnamefont{{M. Jamet et al.}}}, \bibinfo{journal}{Phys. Rev.
  Lett.} \textbf{\bibinfo{volume}{86}}, \bibinfo{pages}{4676}
  (\bibinfo{year}{2001});
%
\bibinfo{author}{\bibnamefont{{M. Jamet et al.}}}, \bibinfo{journal}{Phys. Rev. B}
  \textbf{\bibinfo{volume}{69}}, \bibinfo{pages}{24401} (\bibinfo{year}{2004}).

\bibitem[{\citenamefont{{O. Chubykalo-Fesenko et al.}}(2006)}]{fesenkoetal06preprint}
\bibinfo{author}{\bibnamefont{{O. Chubykalo-Fesenko et al.}}}, \bibinfo{journal}{In preparation}
  (\bibinfo{year}{2006}).

\bibitem[{\citenamefont{{R. Skomski and J.M.D. Coey}}(1999)}]{skocoe99iop}
\bibinfo{author}{\bibnamefont{{R. Skomski and J.M.D. Coey}}},
  \emph{\bibinfo{title}{Permanent Magnetism, Studies in Condensed Matter
  Physics Vol. 1}} (\bibinfo{publisher}{IOP Publishing},
  \bibinfo{address}{London}, \bibinfo{year}{1999}).

\bibitem[{\citenamefont{{K.B. Urquhart et al.}}(1988)}]{urquhartetal88jap}
\bibinfo{author}{\bibnamefont{{K.B. Urquhart, B. Heinrich, J.F. Cochran, A.S.
  Arrott, and Myrtle}}}, \bibinfo{journal}{J. Appl. Phys.}
  \textbf{\bibinfo{volume}{64}}, \bibinfo{pages}{5334} (\bibinfo{year}{1988}).

\bibitem[{\citenamefont{{R. Perzynski and Yu.L.
  Raikher}}(2005)}]{perrai05springer}
\bibinfo{author}{\bibnamefont{{R. Perzynski and Yu.L. Raikher}}}, in
  \emph{\bibinfo{booktitle}{Surface effects in magnetic nanoparticles}}, edited
  by \bibinfo{editor}{\bibfnamefont{D.}~\bibnamefont{Fiorani}}
  (\bibinfo{publisher}{Springer}, \bibinfo{address}{Berlin},
  \bibinfo{year}{2005}), p. \bibinfo{pages}{141}.

\bibitem[{\citenamefont{{Yu. P. Kalmykov}}(2000)}]{kalmykov00prb}
\bibinfo{author}{\bibnamefont{{Yu. P. Kalmykov}}}, \bibinfo{journal}{Phys. Rev.
  B} \textbf{\bibinfo{volume}{61}}, \bibinfo{pages}{6205}
  (\bibinfo{year}{2000}).

\bibitem[{\citenamefont{{A.J. Newell}}(2005{\natexlab{a}})}]{newell05ggg12}
\bibinfo{author}{\bibnamefont{{A.J. Newell}}}, \bibinfo{journal}{Geochem.
  Geophys. Geosyst.} \textbf{\bibinfo{volume}{7}}, \bibinfo{pages}{No. Q03016}
  (\bibinfo{year}{2006}{\natexlab{a}}); 
%
\bibinfo{author}{\bibnamefont{{A.J. Newell}}}, \bibinfo{journal}{Geochem.
  Geophys. Geosyst.} \textbf{\bibinfo{volume}{7}}, \bibinfo{pages}{No. Q03015}
  (\bibinfo{year}{2006}{\natexlab{b}}).

\end{thebibliography}

\end{document}